\documentclass[sigconf]{acmart}
\usepackage{subfiles}
\usepackage{longtable}
\AtBeginDocument{%
  }

\setcopyright{acmlicensed}
\copyrightyear{2018}
\acmYear{2018}
\acmDOI{XXXXXXX.XXXXXXX}
\acmISBN{978-1-4503-XXXX-X/2018/06}

\begin{document}
\begin{sloppypar}

\title{Lifecycle-Aware code generation: Leveraging Software Engineering Phases in LLMs}

\author{Xing Xing}
\authornote{Both authors contributed equally to this research.}
\email{xxing23@m.fudan.edu.cn}
\affiliation{
  \institution{School of Computer Science, Fudan University}
  \city{Shanghai}
  \country{China}
}

\author{Wei Wang}
\authornotemark[1]
\email{wang_wei23@m.fudan.edu.cn}
\affiliation{
  \institution{School of Computer Science, Fudan University}
  \city{Shanghai}
  \country{China}
}

\author{Lipeng Ma}
\email{lpma21@m.fudan.edu.cn}
\affiliation{
  \institution{School of Computer Science, Fudan University}
  \city{Shanghai}
  \country{China}
}

\author{Weidong Yang}
\email{wdyang@fudan.edu.cn}
\authornote{Corresponding author.}
\affiliation{
  \institution{School of Computer Science, Fudan University}
  \city{Shanghai}
  \country{China}
}

\author{Junjie Zheng}
\email{jjzheng22@m.fudan.edu.cn}
\affiliation{
  \institution{School of Computer Science, Fudan University}
  \city{Shanghai}
  \country{China}
}

\renewcommand{\shortauthors}{Trovato et al.}

\begin{abstract}
Recent progress in large language models (LLMs) has advanced automatic code generation, yet most approaches rely on direct, single-step translation from problem descriptions to code, disregarding structured software engineering practices. We introduce a lifecycle-aware framework that systematically incorporates intermediate artifacts such as requirements analysis, state machine modeling, and pseudocode into both the training and inference stages. This design aligns code generation with standard software development phases and enables more structured reasoning. Experiments show that lifecycle-level fine-tuning improves code correctness by up to 75\% over the same model before fine-tuning, with performance gains compounding across intermediate stages. Multi-step inference consistently surpasses single-step generation, demonstrating the effectiveness of intermediate scaffolding. Notably, open-source LLMs, once fine-tuned under our framework, match or slightly outperform models pretrained on code. When applied to DeepSeek-Coder-1.3B, our framework yields relative CodeBLEU improvements of 34.3\%, 20.0\%, 11.2\%, and 22.3\% over ChatGPT-3.5, ChatGPT-4o-mini, DeepSeek-R1, and LLaMA-8B, respectively. Our pipeline also proves robust with up to 80\% less training data, confirming its resilience. Ablation studies further reveal that each intermediate artifact contributes distinctly to final code quality, with state machine modeling yielding the most substantial impact. Our source code and detailed experimental data are available at https://anonymous.4open.science/r/Lifecycle-Aware-3CCB.
\end{abstract}





\received{20 February 2007}
\received[revised]{12 March 2009}
\received[accepted]{5 June 2009}

\maketitle

\section{Introduction}

\begin{figure}[ht]
    \centering
    \includegraphics[width=1\linewidth]{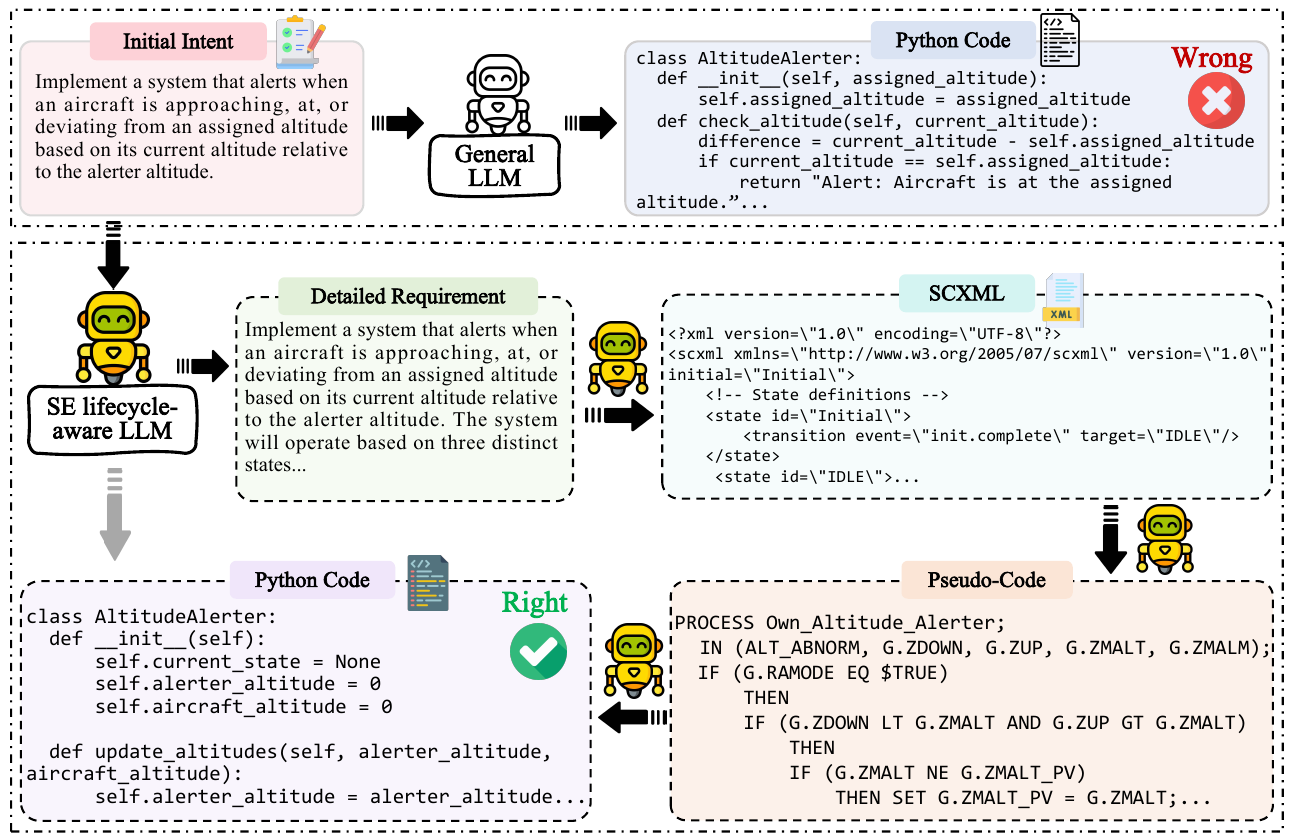}
    \caption{Comparison between single-step code generation and our SE lifecycle-aware multi-step approach.
}
    \label{fig:1}
\end{figure}
Automated code generation has undergone a significant transformation with the advent of large language models (LLMs) such as Codex \cite{chen2021evaluatinglargelanguagemodels}, AlphaCode \cite{li2022competitionlevelcodegenerationalphacode}, and CodeLlama \cite{rozière2024codellamaopenfoundation}. These models have set new records in benchmarks like HumanEval \cite{chen2021evaluating} and APPS \cite{hendrycks2021measuring}, excelling at generating code for specific algorithmic tasks. The traditional LLM approach typically translates natural language problem descriptions directly into executable code in a single step. This method performs well on well-defined tasks with clear functional requirements but often overlooks the structured processes fundamental to robust software development. Benchmarks focusing on functional correctness fail to account for other critical concerns, such as modularity, consistency in interfaces, and adherence to architectural patterns. As Brooks noted \cite{brooks1995mythical}, “The complexity of software is an essential property, not an accidental one,” yet current LLM-driven paradigms tend to neglect this complexity, bypassing established engineering principles. As a result, generated code may suffer from a lack of separation of concerns, fragile error handling, and difficulties in evolving or maintaining the codebase in real-world applications.

In contrast, traditional software engineering (SE) follows a lifecycle approach, starting with requirement refinement and progressing through formal design artifacts before coding begins \cite{royce1970management, somerville2016software}. This yields key deliverables—such as requirement specifications, architectural models, and interface definitions—that support traceability, peer review, and validation \cite{lehtinen2014perceived, perry1989foundations}. However, recent LLM-based methods tend to emphasize downstream activities like unit testing \cite{chen2023evaluating} or retrieval-augmented prompting \cite{zhang2023repocoder}, bypassing early SE stages. As a result, semantic ambiguities and architectural flaws persist, undermining maintainability and trust. The absence of SE lifecycle integration thus remains a major obstacle to adopting LLM-generated software in high-assurance settings.

To bridge this divide, we introduce a code generation framework covered by the software engineering lifecycle, with a specific focus on development phases. The framework integrates requirements analysis, architectural design, and detailed design into the LLM workflow. Our approach decomposes code generation into four verifiable stages: (1) transforming raw requirements into structured specifications, (2) synthesizing SCXML-based ~\cite{barnett2015state} architectural design
, (3) generating language-agnostic pseudocode, and (4) producing executable code. This staged approach enables explicit verification at each stage of software development and ensures traceability from requirements to implementation, a capability missing from the one-step generation paradigm. This rigorous approach prevents the architectural inconsistencies and maintainability issues associated with one-step generation and re-establishes the methodological rigor necessary for highly reliable software development. Fig.~\ref{fig:1} contrasts single-step code generation with our SE lifecycle-driven approach. While single-step methods quickly translate natural language into code, they often lack modularity, error handling, and architectural coherence. In contrast, our staged process—from user intent to final code—enables stepwise validation, improving quality, maintainability, and consistency.

Our evaluation demonstrates that this SE lifecycle-aware approach significantly improves both code correctness and maintainability. Fine-tuning using lifecycle-level intermediate artifacts improves code correctness by up to 75\% compared to the same model before fine-tuning. In comparison to existing commercial models that rely on direct, single-step generation, such as ChatGPT-3.5, ChatGPT-4o-mini, DeepSeek-R1, and LLaMA-8B, our multi-step inference method shows consistent and significant performance improvements. For example, by fine-tuning only DeepSeek-Coder-1.3B under our framework, we achieve relative CodeBLEU improvements of 34.3\% over ChatGPT-3.5, 20.0\% over ChatGPT-4o-mini, 11.2\% over DeepSeek-R1, and 22.3\% over LLaMA-8B, with all baseline models relying solely on single-step generation.. Even when training data is reduced by up to 80\%, our framework remains effective and robust, continuing to outperform these models. This highlights the advantages of incorporating intermediate scaffolding in generating higher-quality, maintainable code.

For building this framework, we created a tuning corpus, which was constructed by extracting state machine (FSM) descriptions and pseudocode pairs from the RTCA/DO-185B certification standard, using the SCXML~\cite{barnett2015state} representation format that complies with the W3C specification, and integrating industrial-grade state machines from the commercial Opennet platform and the open source XState repository. This multi-source data fusion corpus design ensures that each development stage is derived from actual engineering practice, which greatly enhances the traceability and reliability of safety-critical fields such as aerospace and industrial control.

The principal contributions of this work are:

\begin{enumerate}
    \item \textbf{The first LLM framework tailored to the software development phase of the lifecycle}, integrating requirement analysis, W3C-standard architectural design (via SCXML), and detailed design (via pseudocode) as structured intermediate artifacts. This provides a principled alternative to conventional single-step code generation approaches.
    
    \item \textbf{A high-quality, development-phase-aligned FSM dataset} constructed from formal RTCA/DO-185B specifications and real-world industrial implementations (e.g., XState, Simulink, OpenNet), capturing coherent chains across user intent, requirements, SCXML, pseudocode, and final code.
    
    \item \textbf{Comprehensive empirical validation} showing that lifecycle-aware generation significantly improves code quality. When fine-tuned under our framework, DeepSeek-Coder-1.3B achieves CodeBLEU gains of 34.3\%, 20.0\%, 11.2\%, and 22.3\% over single-step baselines ChatGPT-3.5, ChatGPT-4o-mini, DeepSeek-R1, and LLaMA-8B, respectively.

\end{enumerate}

Our evaluation demonstrates that this SE lifecycle-aware approach enhances functional correctness. These findings establish a new paradigm for trustworthy LLM-assisted software development where engineering rigor complements generative capability.

The remainder of this paper is structured as follows: Section 2 contextualizes our work within SE and LLM research. Section 3 details the framework architecture. Section 4 describes our dataset and evaluation protocol. Section 5 presents quantitative findings, while Section 6 explores implications and limitations. Section 7 summarizes contributions and future directions.

\section{Background and Related Work}
\section{SE lifecycle-aware code generation framework}
\begin{figure*}[ht]
  \centering
  \includegraphics[width=\textwidth]{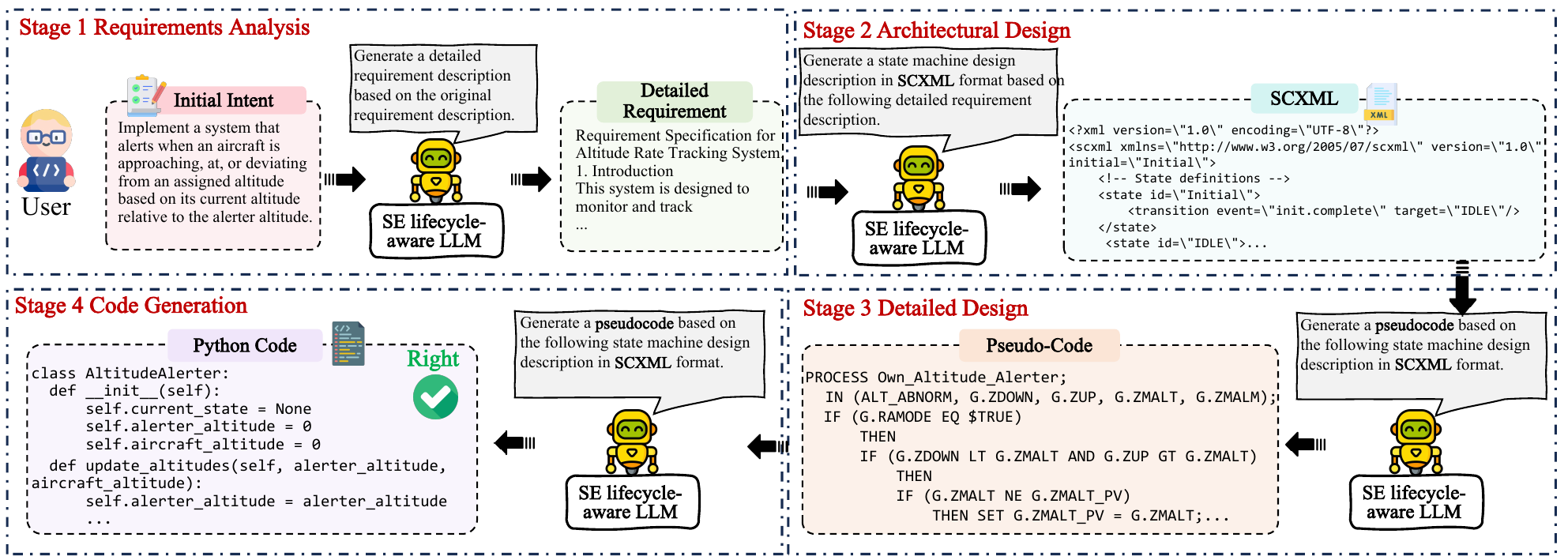}
  \caption{The SE lifecycle-aware code generation framework.}
  \label{fig:2}
\end{figure*}
LLMs have demonstrated strong capabilities in code generation from natural language descriptions, even surpassing human performance on simple tasks~\cite{jimenez2024evaluation}. However, their effectiveness deteriorates on complex problems, with issues such as uncertainty~\cite{ouyang2023llm, donato2025studying}, hallucination~\cite{liu2024exploring}, and repetition~\cite{liu2025code}. Most LLMs focus on isolated functions~\cite{rahman2025large} and still struggle with large-scale, system-level code generation.

\textbf{Prompt Engineering.} This technique involves designing and optimizing instructions to guide LLM outputs~\cite{murr2023testing}, and has been shown to significantly improve result quality~\cite{liu2023pre}. Zhang et al.~\cite{zhang2024renaissance} introduced ILP-style prompts based on Literate Programming~\cite{knuth1984literate} to support large-scale project development. Assogba et al.~\cite{assogbaevaluating} improved long-range dependency handling using annotated function call prompts. M2WF~\cite{wang2025leveraging} employed multi-stage prompting to enhance single-pass codegen in data-free settings. Busch et al.~\cite{busch2025llm} introduced domain-specific natural languages (DSNLs) under the LDE paradigm for task-specific prompting. ChatCoder~\cite{wang2023chatcoder} incorporated multi-angle requirement analysis to improve prompt quality and alignment with user intent.

\textbf{Reinforcement Learning from Feedback.} Feedback-based optimization addresses generation errors and enhances LLM performance. Dutta et al.~\cite{dutta2024applying} used RLAIF to extract AI feedback from GPT-3.5, enabling a 780M model to outperform a 7B baseline. SIT et al.~\cite{sit2024developing} employed error-driven re-prompting to align outputs with user goals. ClarifyGPT~\cite{mu2024clarifygpt} generated clarification questions to refine ambiguous inputs. Ashrafi et al.~\cite{ashrafi2025enhancing} incorporated runtime feedback from code execution to guide multi-agent generation. Xie et al.~\cite{xie2025empowering} leveraged GPT-2 as a planner to improve deep learning code. UnCert-CoT~\cite{zhu2025uncertainty, wei2022chain} triggered reasoning based on uncertainty signals, improving efficiency. CodeLutra~\cite{tao2024codelutra} used iterative self-generated error correction for fine-tuning.

\textbf{Multi-stage Generation Mechanisms.} To better handle multi-module systems, recent studies introduce intermediate representations. Zuo et al.~\cite{zuo2025complexvcoder} proposed GIR to improve NL-to-Verilog translation. CoDes~\cite{vijayaraghavan2024chain} employed formulate-refine-execute planning for VHDL generation. Murphy et al.~\cite{murphy2024combining} integrated TSL specifications to improve correctness. Zhao et al.~\cite{zhao2024empowering} synthesized HDL summaries for fine-tuning. Yaacov et al.~\cite{yaacov2024boosting} introduced Behavioral Programming as an intermediate to enhance requirement coverage.

\textbf{Multi-agent Frameworks.} Single LLMs are often insufficient for complex tasks, prompting multi-agent approaches. Almorsi et al.~\cite{almorsi2025guided} used problem-tree decomposition to outperform llama3.1-8B by 23.79\%. Agents4PLC~\cite{liu2024agents4plc} used modular agents for NL-to-PLC generation. FlowGen~\cite{lin2024llm} embedded LLMs into standard software processes like Waterfall and TDD. EnsLLM~\cite{mahmud2025enhancing} applied ensemble voting with CodeBLEU~\cite{ren2020codebleu} and CrossHair metrics to improve robustness. CodeCoR~\cite{pan2025codecor} and AutoP2C~\cite{lin2025autop2c} further boosted multi-agent performance through collaboration evaluation and multimodal input processing.

\subsection{Framework Overview}
The proposed SE lifecycle-aware code generation framework redefines the paradigm for LLM-assisted software development by integrating essential software engineering lifecycle phases into a unified, trainable architecture. As depicted in Fig.~\ref{fig:2}, the framework decomposes code generation into four sequential stages:

\textbf{requirement analysis}: 
Given a high-level and often ambiguous user intent, the LLM performs contextual reasoning to infer a structured and detailed requirement specification. This intermediate artifact reduces semantic ambiguity and provides a formalized basis for the subsequent design phase.

\textbf{architectural design}: 
Building upon the requirements analysis, the LLM constructs a formal behavioral model of the system using SCXML—a machine-readable representation of a finite state machine.
This state machine captures control logic, transitions, and system states in a declarative form, serving as a bridge between functional requirements and executable logic.

\textbf{detailed design}:
The LLM translates the declarative SCXML into structured pseudocode that captures algorithmic logic and control flow.
Serving as a language-agnostic intermediate representation, the pseudocode enhances readability, supports traceability, and enables seamless integration with downstream LLM-based code generation systems.
This abstraction is particularly valuable for generalizing the framework to support multiple target languages.

\textbf{code generation}: 
Based on the structured pseudocode, LLM generates executable code that meets both functional requirements and design constraints.
This phase can leverage the power of LLM to ensure syntactic correctness and language-specific optimizations.
The modular nature of this phase also allows replacement or extension with language-specific backends, ensuring that it can adapt to a variety of deployment environments.

The key feature of the framework is contextual inheritance across stages: the output of each stage becomes the structured input to the next. This design mirrors the development phase within the software lifecycle, ensuring continuity of intent from requirements to implementation. Each transition preserves semantic information, enhancing coherence and consistency throughout the engineering process.
Moreover, the framework provides flexible inference capabilities, supporting both sequential end-to-end generation from raw requirements to final code implementation, as well as targeted stage-specific generation that allows isolated processing of any intermediate artifact to facilitate partial refinement or human-in-the-loop collaborative workflows.

\subsection{Dataset Construction}
\begin{figure}[ht]
    \centering
    \includegraphics[width=1\linewidth]{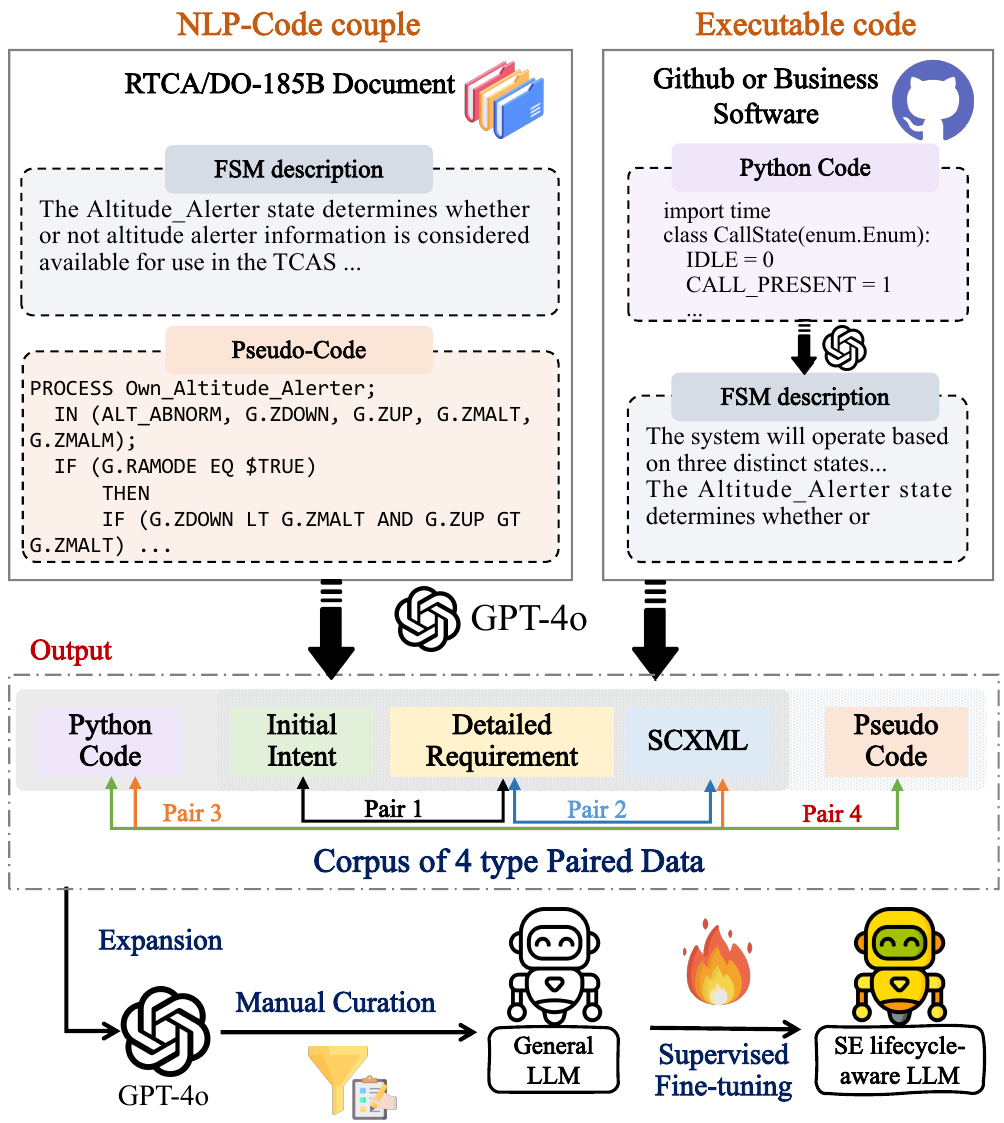}
    \caption{Overview of the dataset construction pipeline integrating specification-derived and implementation-derived artifacts.}
    \label{fig:3}
\end{figure}

Fig.~\ref{fig:3} shows the pipeline for constructing a dataset of aligned artifacts across four key software development stages. The goal is to form semantically consistent transformation pairs linking intent, intermediate design artifacts, and final code. We started by manually extracting and structuring FSM descriptions and corresponding pseudocode from the RTCA/DO-185B~\cite{rtca2009do185b} standard. This required interpreting formal mappings from state logic to control flow, where each FSM includes detailed states, transitions, guard conditions, and normative pseudocode capturing intended behaviors. 

To enrich the dataset with realistic requirements and formal encodings, we used GPT-4o to generate additional representations for each FSM. For every FSM, the model produced a plain-language intent narrative and a structured requirement analysis covering functional constraints, responsibilities, modes, I/O relations, and safety or timing conditions. This was then translated into SCXML, creating a trace from natural-language intent to machine-readable specification. Finally, GPT-4o converted the pseudocode into executable Python code, forming an end-to-end artifact chain. All prompt templates are listed in Appendix. In parallel, we collected real-world Python implementations from three representative sources: \textbf{XState} (via official documentation~\footnote{\url{https://stately.ai/docs/examples}} and GitHub~\footnote{\url{https://github.com/statelyai/xstate}}), \textbf{Simulink} (via demo models~\footnote{\url{https://ww2.mathworks.cn/help/simulink/examples.html}} and GitHub projects~\footnote{\url{https://github.com/mathworks/MATLAB-Simulink-Challenge-Project-Hub}}), and \textbf{OPNET} (via a licensed installation~\footnote{\url{https://support.riverbed.com/content/support/software/opnet-model/modeler.html}}).

We manually filtered these to retain well-documented implementations of state machines. Using GPT-4o, we reconstructed pseudocode and inferred FSM descriptions from the code, then generated requirement analysis, SCXML, and pseudocode via the same prompt sequence for consistency. To broaden coverage beyond avionics, networking, and software domains, we applied GPT-4o to evolve existing examples, increasing diversity. Human annotators screened these evolved samples and retained only high-quality outputs.

The dataset was organized into four adjacent pairs: intent-to-requirement, requirement-to-SCXML, SCXML-to-pseudocode, and pseudocode-to-code. Each pair was manually reviewed to ensure semantic fidelity. Ambiguous or inconsistent examples were discarded to maintain quality. The final dataset combines normative standard-derived and real-world reconstructed instances, supporting research on traceability and generation from informal to executable artifacts. All prompts and example queries are provided in Appendix for reproducibility.

\subsection{Training Strategy}
To train our SE lifecycle-aware code generation framework, we adopt an end-to-end fine-tuning approach on a unified dataset. Each training instance in our corpus includes a full sequence of artifacts across all four stages: the inital user intent, the detailed requirement analysis, the SCXML-based state machine design, the synthesized pseudocode, and the executable code. These complete data tuples are collected and aligned from domain-specific sources, including aviation standards, commercial software documentation, and open-source repositories. To efficiently adapt LLMs to our task while minimizing computational overhead, we apply the Low-Rank Adaptation (LoRA) technique~\cite{hu2022lora} during fine-tuning. This parameter-efficient method enables us to update a small subset of trainable weights while keeping the vast majority of the base model frozen, allowing fast adaptation to domain-specific patterns with limited resource cost. By fine-tuning the base LLM on this integrated dataset, the model learns the hierarchical mapping across all stages in a holistic manner. Unlike modular or stage-wise fine-tuning strategies, our approach encourages the model to internalize the structural dependencies between phases and to reason coherently from user intent to final implementation. This unified training design simplifies deployment and improves consistency across the generation pipeline.

\subsection{Requirements Analysis}
 The main goal of requirement analysis phase is to transform high-level, usually concise user intents into precise and structured specifications that can be reliably used by downstream stages. The input to this stage is a free-form, informal user intent written in natural language. The expected output is a structured requirement document that comprises the following elements: functional requirements, non-functional requirements, interface specifications, safety and timing constraints, and acceptance criteria.

The fine-tuning data for this stage consists of paired examples, each comprising an informal user intent and its corresponding structured requirement specification. At inference phase, the model extracts intial intents, identifies key entities and operations, infers missing details such as edge cases and data constraints, and outputs structured, hierarchical specifications (including markdown-style section headings, bullet lists, and code blocks) via a lightweight prompt template that enforces a consistent format.

The explicit generation of structured requirements artifacts helps reduce semantic ambiguity while providing multiple benefits: it establishes comprehensive traceability by associating all design decisions with concrete requirements specifications; it significantly reduces errors by preventing downstream misunderstandings through precisely defined acceptance criteria; and it enables an efficient human-computer interaction workflow by allowing engineers to review and modify the generated specifications. Experimental results validate the effectiveness of the approach and demonstrate its robustness as the basis for an LLM-driven code generation system.

\subsection{Architectural Design}
The architectural design stage formalizes system behavior using state machine representations encoded in XML. State machines provide a structured and precise way to model behaviors driven by discrete events and conditional transitions. Their modularity and formal semantics make them effective for reactive systems, embedded controllers, user interfaces, and communication protocols~\cite{harel1987statecharts}, while also applicable to general-purpose software involving control flow and execution logic. As a result, they serve as powerful intermediate abstractions in code generation pipelines—human-interpretable, machine-verifiable, and conducive to traceability.

We adopt the State Chart XML (SCXML) standard defined by the W3C~\cite{barnett2015state} to encode these state machines in a machine-readable and schema-compliant format. SCXML supports event-driven transitions, hierarchical states, and parallel composition. Compared to informal behavioral descriptions, SCXML offers well-defined semantics, structural rigor, and compatibility with validation tools, making it well-suited for downstream reasoning and synthesis in LLM-driven workflows. The input to this stage is a structured requirement specification that includes functional descriptions, event conditions, and expected behaviors. The output is an SCXML document conforming to the W3C state machine specification, which defines states, transitions, events, actions, and, where applicable, a hierarchical structure.

The fine-tuning data for this stage consists of paired examples mapping detailed requirements to their corresponding SCXML representations. During inference phase, the model identifies distinct control states, relevant triggering events, and transition conditions, and outputs a schema-compliant SCXML. Generated artifacts are directly parsable and can be validated using standard SCXML engines.

This stage formalizes the behavioral semantics of the system before procedural code is generated. By grounding the design in a standard executable model, it improves transparency, enables automated state-based analysis, and establishes a strong logical foundation for downstream detailed design. In practice, the use of SCXML improves error isolation and contributes to better traceability from requirements to implementation.

\subsection{Detailed Design}
In this stage, the pseudocode serves as the primary artifact of the detailed design phase. A language independent intermediate representation is introduced between the FSM design based on SCXML and the executable code. Based on the state machine, LLM synthesizes pseudocode that manipulates the control logic and structure captured in SCXML. In this sense, pseudocode acts as an executable projection of the state machine, transforming declarative transformations into imperative logic. The input to this stage is an SCXML document representing the system’s control logic, including defined states, transitions, and event-driven behaviors. The output is a structured pseudocode artifact that imperatively realizes the state machine logic, translating the declarative SCXML specification into a sequential, human-readable form that reflects its execution semantics.

Based on the SCXML artifact generated in the previous stage, the LLM synthesizes structured pseudocode that reflects the control logic, event handlers, and transition conditions encoded in the state machine. This transformation projects the declarative semantics of SCXML into imperative constructs such as conditionals, loops, and function calls—yielding a representation that is human-readable, algorithmically complete, and structurally aligned with conventional programming logic.

Using pseudocode as an intermediate abstraction facilitates downstream integration with existing LLMs specifically designed for code generation. Many state-of-the-art LLMs, such as Codex and CodeLlama, have demonstrated powerful performance in transforming structured pseudocode into executable code across various programming languages. By separating design logic from specific syntax in the early stages of the generation process, our framework gains scalability and modularity, enabling seamless support for future language goals and specialized generation models. 

\subsection{code generation}
The code generation phase corresponds to the implementation phase of the SE lifecycle. In this phase, the framework converts the structured, language-independent pseudocode into executable code in our current implementation, Python. This step completes the transition from abstract design to concrete implementation while ensuring that the control flow, behavioral logic, and structural integrity of the earlier phases are faithfully preserved. The input to this stage is structured pseudocode that expresses the system’s logic in an imperative, language-neutral form. The output is executable Python code that faithfully implements the behavior defined in the pseudocode.

Unlike traditional approaches that directly rely on pre-trained code generation LLMs, we fine-tune a foundation LLM specifically for the task of pseudocode-to-Python translation. This supervised adaptation enables the model to learn precise alignment patterns between abstract logic and idiomatic Python syntax.

Although our current focus is on Python, the framework is model-agnostic and modular by design. The pseudocode representation decouples design logic from language-specific syntax, enabling integration with various downstream code generators or multilingual backends. Thus, the framework supports seamless extension to other programming languages (e.g., Java, C++) or plug-in replacement of the code generation module with specialized LLMs such as Codex\cite{rozière2024codellamaopenfoundation} or CodeLlama~\cite{roziere2023code}, depending on future deployment needs. 

\section{Experimental Setup}

\subsection{Dataset Construction}
Our dataset comprises four paired artifacts reflecting sequential SE stages. We first extracted FSM descriptions and their corresponding pseudocode from the RTCA/DO-185B standard, which provides formal mappings from state machine specifications to structured pseudocode. Using GPT-4o, we inferred detailed requirements analysis, human intent narratives, and SCXML representations, which serve as standardized XML-based encodings of state machines, based on each FSM description.

Additionally, we sampled two open-source frameworks (XState, Simulink) and one commercial tool (OpenNet) from GitHub repositories. From each, we manually filtered Python implementations of state machine logic. We then applied GPT-4o to reverse-engineer pseudocode from the Python sources and used GPT-4o to reconstruct corresponding FSM descriptions. For these reconstructed FSM descriptions, GPT-4o generated detailed requirement analysis, human intent narratives, and SCXML representations following the same procedure as applied to the RTCA/DO-185B data. Additionally, to enhance dataset diversity, we used GPT-4o to expand upon the existing seed data, aiming for broad coverage across various fields. Human annotators then screened the generated data, retaining high-quality examples for inclusion in the dataset.

Through this pipeline, we obtained four aligned artifact pairs:
\begin{enumerate}
  \item Human intent $\rightarrow$ Detailed requirement analysis
  \item Detailed requirement analysis $\rightarrow$ SCXML
  \item SCXML $\rightarrow$ Pseudocode
  \item Pseudocode $\rightarrow$ Python code
\end{enumerate}
All pairs underwent manual quality screening to retain high-fidelity inference chains, ensuring the integrity of the four-stage rationale for downstream fine-tuning.

\subsection{Evaluation Metrics}
Generation quality at each stage is assessed using established automatic metrics. Specifically, for requirements analysis, architectural design, and detailed design, we employ Exact Match (EM), BLEU~\cite{papineni2002bleu}, and ROUGE-L~\cite{lin2004rouge} to capture both syntactic fidelity and content overlap against reference artifacts. For the architectural design stage, we additionally apply TF-IDF~\cite{salton1988term} to assess the keyword relevance and term coverage between the generated SCXML files and the ground-truth specifications, which are typically semi-structured and emphasize critical transitions and event-driven behaviors.

To evaluate the final Python code output, we adopt the CodeBLEU metric~\cite{ren2020codebleu}, which extends traditional BLEU by incorporating code-specific features such as syntax structure, data flow analysis, and semantic similarity. All metric scores are computed on a held-out test subset representing 20\% of the manually screened artifact pairs, ensuring no data leakage from the fine-tuning process. This evaluation protocol provides a systematic and multi-perspective means to quantify improvements introduced at each intermediate stage and the overall pipeline.

\subsection{Baseline Models and Backbone Selection}
Our experiments leverage both language-specialized and code-specialized foundation models. The language-specialized cohort comprises Qwen-2.5-Instruction\cite{team2024qwen2} with 1.5 billion and 7 billion parameters, selected for their demonstrated competence in general natural language understanding and generation. In parallel, the code-specialized cohort includes DeepSeek-Coder-1.3B-Base\cite{guo2024deepseek} and DeepSeek-Coder-6.7B-Base\cite{guo2024deepseek}, which are optimized for program synthesis tasks.

To contextualize the gains from our integrated SE lifecycle-aware fine-tuning methodology, we benchmark these fine-tuned models against four state-of-the-art baselines without additional training: GPT-4o-Mini\cite{achiam2023gpt}, GPT-3.5-Turbo\cite{brown2020language}, DeepSeek-R1\cite{guo2025deepseek}, and LLaMA-8B\cite{touvron2023llama}. By comparing performance across this diverse set of models, we isolate the contributions of pipeline structuring and multi-phase learning from raw model capacity.

\section{Results and Analysis }
We formulate five research questions to evaluate our SE lifecycle-
aware code generation framework
\begin{table*}[htbp]
  \centering
  \caption{Stage-wise performance of base and fine-tuned models across the SE lifecycle, measured by EM, BLEU, ROUGE, TF-IDF, and CodeBLEU}
  \label{tab:1}
    \begin{tabular}{c|ccc|ccc|ccc|ccc}
    \toprule
    \multicolumn{1}{c}{} & \multicolumn{3}{c}{Requirements Analysis} & \multicolumn{3}{c}{Architectural Design} & \multicolumn{3}{c}{Detailed Design} & \multicolumn{3}{c}{Code Generation} \\
    \midrule
          & em    & bleu  & rouge & em    & bleu  & tf-idf & em    & bleu  & rouge & em    & bleu  &  codebleu \\
    \midrule
    DSK-C-1.3B & 0.1226  & 0.0033  & 0.1238  & 0.0240  & 0.0019  & 0.1058  & 0.0329  & 0.0006  & 0.0467  & 0.0467  & 0.0006  & 0.1699  \\
    DSK-C-1.3B-F & 0.3259  & 0.0275  & 0.1877  & 0.2449  & 0.3115  & 0.5010  & \textbf{0.1482 } & \textbf{0.0040 } & \textbf{0.1203 } & \textbf{0.1791 } & \textbf{0.0190 } & 0.2971  \\
    DSK-C-6.7B & 0.1925  & 0.0100  & 0.1492  & 0.0401  & 0.0159  & 0.1179  & 0.0348  & 0.0007  & 0.0624  & 0.0479  & 0.0006  & 0.1633  \\
    DSK-C-6.7B-F & 0.3753  & 0.0432  & 0.2137  & 0.3307  & 0.5276  & 0.7383  & 0.1773  & 0.0056  & 0.1538  & 0.2130  & 0.0315  & 0.2863  \\
    Qwen-1.5B & 0.2936  & 0.0233  & 0.1148  & 0.2102  & 0.3037  & 0.6284  & 0.0842  & 0.0016  & 0.1227  & 0.1183  & 0.0062  & 0.2178  \\
    Qwen-1.5B-F & 0.3707  & 0.0370  & 0.1909  & \textbf{0.3522 } & 0.4435  & 0.6579  & 0.1801  & 0.0056  & 0.1620  & 0.2014  & 0.0281  & 0.2703  \\
    Qwen-7B & 0.3007  & 0.0321  & 0.1219  & 0.2434  & 0.3329  & 0.7214  & 0.0969  & 0.0018  & 0.0901  & 0.1350  & 0.0090  & 0.2694  \\
    Qwen-7B-F & \textbf{0.4185 } & \textbf{0.0582 } & \textbf{0.2606 } & 0.3457  & \textbf{0.5467 } & 0.7745  & \textbf{0.2055 } & \textbf{0.0086 } & \textbf{0.1965 } & \textbf{0.2453 } & \textbf{0.0463 } & 0.2923  \\
    GPT-4o-mini & 0.3302  & 0.0399  & 0.1410  & 0.2571  & 0.3208  & 0.7395  & 0.0989  & 0.0019  & 0.0988  & 0.1584  & 0.0114  & 0.2740  \\
    GPT-3.5 & 0.3325  & 0.0414  & 0.2216  & 0.2989  & 0.5774  & \textbf{0.8305 } & 0.1177  & 0.0030  & 0.1923  & 0.0930  & 0.0018  & 0.1374  \\
    DSK-R1 & 0.2777  & 0.0128  & 0.1096  & 0.2119  & 0.2083  & 0.6949  & 0.0906  & 0.0014  & 0.0838  & 0.1656  & 0.0133  & \textbf{0.3087 } \\
    LLaMA-8B & 0.2849  & 0.0317  & 0.1379  & 0.2668  & 0.3925  & 0.7499  & 0.1062  & 0.0022  & 0.1136  & 0.1558  & 0.0108  & 0.2982  \\
    \bottomrule
    \end{tabular}%
  \label{tab:addlabel}%
\end{table*}%

\subsection{RQ1: How does integrated SE-lifecycle fine-tuning influence performance in different phases of code generation?}

We evaluate whether multi-stage supervision based on the SE lifecycle improves code generation quality by comparing each base model with its fine-tuned counterpart of the same architecture and scale. As shown in Table~\ref{tab:1}, SE-lifecycle fine-tuning significantly enhances the correctness of both intermediate and final code across all evaluated models. For example, the CodeBLEU score of DeepSeek-coder-1.3B (DSK-C-1.3B) improves from 0.1699 to 0.2971 (+74.9\%), while that of DeepSeek-coder-6.7B (DSK-C-6.7B) rises from 0.1633 to 0.2863 (+75.3\%). These results show that explicitly modeling SE artifacts such as requirements, designs, and pseudocode can significantly enhance the LLM's landing ability and output fidelity. 

The performance gain is not limited to code-centric models. The general-purpose Qwen2.5-7B (Qwen-7B) improves by 8.5\% in CodeBLEU (from 0.2694 to 0.2923), indicating that SE lifecycle supervision is broadly effective across model families. Notably, smaller fine-tuned models often outperform larger untuned ones. For example, fine-tuned DSK-C-1.3B (DSK-C-1.3B-F) achieves a CodeBLEU of 0.2971, surpassing ChatGPT-3.5-turbo (0.1374) and ChatGPT-4o-mini (0.2740), and nearly matching DeepSeek-R1 (DSK-R1, 0.3087), despite its smaller size.These results demonstrate the effectiveness of process-aware supervision through fine-tuning and highlight that specialized models, even at smaller scales, can perform competitively or even better than much larger general-purpose or non-specialized models when guided by high-quality task-specific supervision.

A fine-grained breakdown shows that improvements accumulate across the pipeline. In the requirements analysis phase, DSK-C-1.3B improves EM from 0.1226 to 0.3259 (+95.1\%), generating more complete specifications. This propagates to the design phase, where SCXML TF-IDF similarity increases from 0.1058 to 0.5010 (+374\%), reflecting better modeling of control logic. In detailed design, ROUGE improves 158\% (from 0.0467 to 0.1203), contributing to the final-stage CodeBLEU gain of +74.9\%. These results support our hypothesis: refining early-stage artifacts provides strong inductive bias for later stages, enabling more accurate and verifiable code generation.

Notably, our fine-tuned models also rival or outperform high-performing API-based LLMs. Tuned DSK-C-1.3B outperforms ChatGPT-3.5-turbo and ChatGPT-4o-mini, and nearly matches DSK-R1 (CodeBLEU = 0.3087), highlighting that SE lifecycle integration can narrow the gap between open-source and proprietary models. These findings confirm the practical value of our approach and show that SE lifecycle grounding not only improves statistical performance but also enhances real-world applicability.

\subsection{RQ2: How does multi-step inference compare to single-step inference in output quality?}

\begin{table*}[htbp]
  \centering
  \caption{Stage-wise performance gap (single-step versus multi-step) across EM, BLEU, ROUGE, TF-IDF, and CodeBLEU metrics.}
  \label{tab:2}
    \begin{tabular}{c|ccc|ccc|ccc}
    \toprule
    \multicolumn{1}{c}{} & \multicolumn{3}{c}{Architectural Design} & \multicolumn{3}{c}{Detailed Design} & \multicolumn{3}{c}{Code Generation} \\
    \midrule
          & em    & bleu  & tf-idf &  em   & bleu  & rouge & em    & bleu  & codebleu \\
    \midrule
    DSK-C-1.3B & 0.0168  & 0.0004  & \textbf{-0.0020 } & 0.0164  & 0.0003  & 0.0164  & 0.0257  & 0.0006  & \textbf{-0.0037 } \\
    DSK-C-1.3B-F & \textbf{-0.0064 } & \textbf{-0.0059 } & 0.0034  & \textbf{-0.0752 } & \textbf{-0.0025 } & \textbf{-0.0335 } & \textbf{-0.0096 } & \textbf{-0.0053 } & \textbf{-0.0560 } \\
    DSK-C-6.7B & 0.0394  & 0.0347  & 0.0315  & 0.0282  & 0.0003  & 0.0162  & 0.0377  & 0.0013  & 0.0100  \\
    DSK-C-6.7B-F & 0.0051  & 0.0072  & 0.0340  & \textbf{-0.0371 } & \textbf{-0.0014 } & \textbf{-0.0149 } & \textbf{-0.0167 } & \textbf{-0.0060 } & \textbf{-0.0484 } \\
    Qwen-1.5B  & \textbf{-0.0224 } & 0.0845  & \textbf{-0.0822 } & 0.0247  & 0.0005  & 0.0132  & 0.0215  & 0.0020  & 0.0072  \\
    Qwen-1.5B-F  & \textbf{-0.0028 } & 0.0395  & \textbf{-0.0208 } & \textbf{-0.0521 } & \textbf{-0.0023 } & \textbf{-0.0072 } & \textbf{-0.0267 } & \textbf{-0.0095 } & \textbf{-0.0350 } \\
    Qwen-7B & 0.0403  & 0.0936  & \textbf{-0.0104 } & 0.0182  & 0.0005  & 0.0111  & 0.0362  & 0.0053  & \textbf{-0.0227 } \\
   Qwen-7B-F  & 0.0104  & 0.0298  & 0.0371  & \textbf{-0.0029 } & \textbf{-0.0007 } & 0.0320  & \textbf{-0.0122 } & \textbf{-0.0079 } & \textbf{-0.0406 } \\
   GPT-4o-mini & 0.0277  & 0.0861  & 0.0026  & 0.0331  & 0.0010  & 0.0230  & 0.0154  & 0.0020  & \textbf{-0.0264 } \\
    GPT-3.5 & 0.0047  & 0.0237  & \textbf{-0.0101 } & \textbf{-0.0085 } & \textbf{-0.0010 } & \textbf{-0.0180 } & 0.0318  & 0.0022  & 0.0839  \\
    DSK-R1 & 0.0798  & 0.1809  & 0.0662  & 0.0329  & 0.0017  & 0.0252  & 0.0253  & 0.0055  & \textbf{-0.0415 } \\
    LLaMA-8B & 0.0277  & 0.0027  & \textbf{-0.0796 } & 0.0025  & \textbf{-0.0002 } & 0.0109  & 0.0089  & \textbf{-0.0001 } & \textbf{-0.0553 } \\
    \bottomrule
    \end{tabular}%
  \label{tab:addlabel}%
\end{table*}%

To investigate whether decomposing code generation into intermediate steps improves output quality, we compare multi-step and single-step inference across three stages: architectural design (AD), detailed design (DD), and code generation (CG). The requirements analysis (RA) stage is excluded, as both inference modes yield same results. In single-step inference, models map user intent directly to artifacts (e.g., SCXML, pseudocode, or code). In contrast, multi-step inference uses structured intermediates like detailed requirements, SCXML, and pseudocode as scaffolding. Critically, all fine-tuned models in this comparison were specifically adapted to either multi-step or single-step datasets, thereby isolating the influence of reasoning structures from pre-training or model scale effects. 

As shown in Table~\ref{tab:2}, multi-step inference consistently outperforms single-step inference across the majority of models and stages, especially in the code generation phase. For instance, the DSK-C-1.3B-F exhibits a decrease of $0.0560$ in CodeBLEU and $0.0096$ in EM when using single-step inference, indicating degraded syntactic and semantic correctness in directly generated code. Similar trends are observed in qwen-1.5B-F and qwen-7B-F, where CodeBLEU differences are $0.0350$ and $0.0406$ respectively. These declines suggest that bypassing intermediate artifacts leads to error accumulation and weaker inductive structure in final outputs. 

Multi-step inference also shows clear advantages in intermediate stages such as detailed design. Most fine-tuned models show negative deltas in ROUGE and BLEU metrics. For example, qwen-1.5B-F drops by $0.0072$ in ROUGE and $0.0023$ in BLEU. The DSK-C-6.7B-F model loses $0.0149$ in ROUGE and $0.0014$ in BLEU, reinforcing that directly generating pseudocode from raw requirements omits crucial semantic transformations provided by design artifacts like SCXML. 

Even large models like ChatGPT-3.5, ChatGPT-4o-mini, and DeepSeek-R1 show noticeable performance drops in single-step settings. DeepSeek-R1, for instance, suffers a $0.0415$ CodeBLEU loss in code generation under single-step inference. This indicates that model size alone cannot compensate for missing intermediate structure.

On the other hand, models fine-tuned with single-step data show only modest improvements or neutral performance when used in a single-step inference pipeline. However, in aggregate, models trained and evaluated in a multi-step fashion consistently outperform their single-step counterparts. This is particularly evident in aggregate metrics such as CodeBLEU and TF-IDF similarity for state machine outputs. For instance, the DSK-C-6.7B-F model shows a +0.0340 gain in TF-IDF when using multi-step inference. 

Critically, open-source LLMs fine-tuned with our framework match or surpass the performance of leading commercial code models. For instance, DeepSeek-Coder-1.3B reaches a CodeBLEU of 0.2971, surpassing ChatGPT-3.5 (0.2213, +34.3

These results strongly support the hypothesis that multi-step inference provides valuable structural inductive bias, improving compositional generalization and semantic fidelity across all stages of the SE lifecycle. By leveraging intermediate representations such as SCXML and pseudocode, LLMs are better able to reason hierarchically, produce more faithful outputs, and avoid semantic drift in generation tasks.

\subsection{RQ3: How do general LLMs and code-pretrained LLMs differ in performance after fine-tuning?}
\begin{figure}[h]
  \centering
  \includegraphics[width=\linewidth]{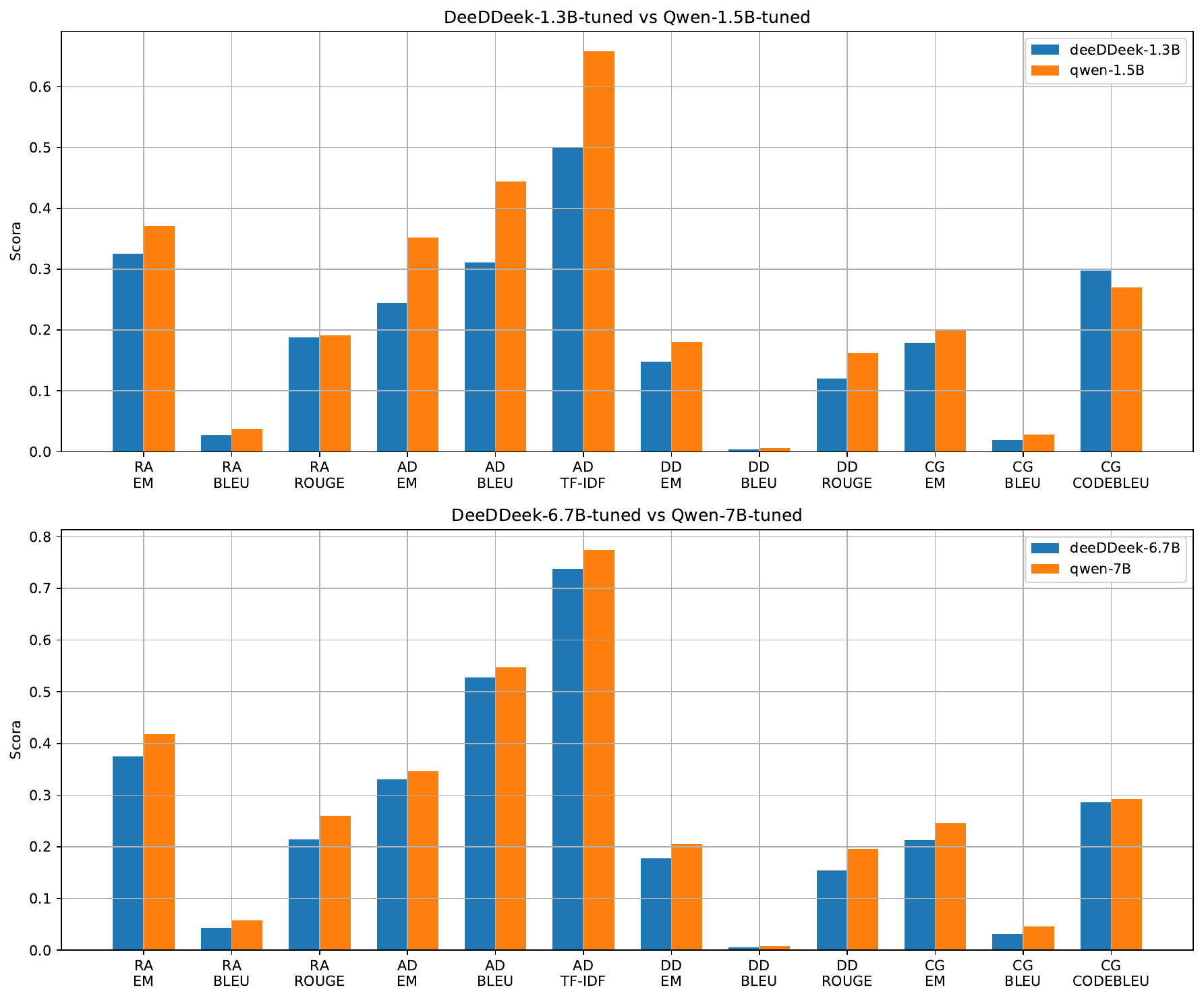}
  \caption{Comparison of general-purpose (Qwen2.5) and code-pretrained (DeepSeek-Coder) LLMs after SE-lifecycle fine-tuning across SE stages (RA: requirement analysis; AD: architectural design; DD: detailed design; CG: code generation). 
}
  \label{fig:4}
\end{figure}
To compare general-purpose and code-pretrained LLMs under fine-tuning, we evaluate Qwen2.5 models (Qwen-1.5B and Qwen-7B) against similarly sized Deepseek-Coder models (DSK-C-1.3B and DSK-C-6.7B). This comparison allows us to isolate the effect of pretraining priors (natural language vs. code-centric) on the models' capacity to understand SE processes and produce high-quality code artifacts after supervised fine-tuning.

As shown in Fig.~\ref{fig:4}, Qwen2.5 models often match or outperform Deepseek-Coder after identical fine-tuning. In the requirements analysis (RA) phase, Qwen-1.5B-F achieves higher scores than DSK-C-1.3B in EM (0.3707 vs. 0.3259), BLEU (0.0370 vs. 0.0275), and ROUGE (0.1909 vs. 0.1877). Similar trends appear in architectural design (AD), where Qwen-1.5B-F again leads in EM (0.3522 vs. 0.2449), BLEU (0.4435 vs. 0.3115), and TF-IDF (0.6579 vs. 0.5010). These results show that language-pretrained models can effectively internalize formal specifications and control logic when provided with structured supervision.

Qwen-7B-F continues this trend at a larger scale. In detailed design (DD), it outperforms DSK-C-6.7B-F in ROUGE (0.1965 vs. 0.1538) and BLEU (0.0086 vs. 0.0056). In code generation (CG), Qwen-7B-F achieves higher CodeBLEU (0.2923 vs. 0.2863), EM (0.2453 vs. 0.2130), and BLEU (0.0463 vs. 0.0315). These results indicate that despite their lack of code-specific pretraining, general LLMs with strong language modeling capabilities and sufficient capacity are highly effective in downstream code generation when guided through semantically grounded, multi-step supervision.

Notably,  Qwen2.5’s advantage is more pronounced in earlier, language-heavy stages (RA, AD), indicating that general LLMs benefit from richer linguistic priors when handling abstract or ambiguous inputs. In contrast,  Deepseek-Coder shows more stable performance in later stages (e.g., code generation), reflecting stronger syntactic and code-level fluency. However, after fine-tuning, these differences are small, and Qwen2.5 often delivers comparable or better overall performance.

These findings suggest that general-purpose LLMs are competitive—even advantageous—for software engineering tasks when augmented with SE lifecycle fine-tuning pipelines. Lifecycle-aware supervision not only compensates for the lack of code-centric pretraining but can elevate general models to match or surpass specialized models in both intermediate representation and final code generation quality. This reinforces the value of our methodology as a scalable and model-agnostic approach to program synthesis.

\subsection{RQ4: How does training data size affect pipeline robustness?} 

\begin{table*}[htbp]
  \centering
  \caption{Performance of the Qwen2.5-7B base model after fine-tuning on 20-100\% of the training data, evaluated on an identical held-out test set across the requirements analysis, architectural design, detailed design, and code generation stages.}
  \label{tab:3}
    \begin{tabular}{c|rrr|rrr|rrr|rrr}
    \toprule
    \multicolumn{1}{c}{} & \multicolumn{3}{c}{Requirements Analysis} & \multicolumn{3}{c}{Architectural Design} & \multicolumn{3}{c}{Detailed Design} & \multicolumn{3}{c}{Code Generation} \\
    \midrule
          & \multicolumn{1}{c}{em} & \multicolumn{1}{c}{bleu} & \multicolumn{1}{c|}{rouge} & \multicolumn{1}{c}{em} & \multicolumn{1}{c}{bleu} & \multicolumn{1}{c|}{tf-idf} & \multicolumn{1}{c}{em} & \multicolumn{1}{c}{bleu} & \multicolumn{1}{c|}{rouge} & \multicolumn{1}{c}{em} & \multicolumn{1}{c}{bleu} & \multicolumn{1}{c}{code bleu} \\
    \midrule
    100\% & 0.4185  & 0.0582  & 0.2606  & 0.3457  & 0.5467  & 0.7745  & 0.2055  & 0.0086  & 0.1965  & 0.2453  & 0.0463  & 0.2923  \\
    80\%  & 0.4260  & 0.0642  & 0.2474  & 0.3688  & 0.5766  & 0.7853  & 0.2095  & 0.0093  & 0.2089  & 0.2448  & 0.0443  & 0.2833  \\
    60\%  & 0.4251  & 0.0635  & 0.2462  & 0.3675  & 0.5589  & 0.7829  & 0.2158  & 0.0088  & 0.1966  & 0.2401  & 0.0450  & 0.2905  \\
    40\%  & 0.4170  & 0.0599  & 0.2596  & 0.3660  & 0.5751  & 0.8073  & 0.2092  & 0.0088  & 0.1872  & 0.2427  & 0.0459  & 0.2859  \\
    20\%  & 0.4139  & 0.0602  & 0.2384  & 0.3565  & 0.5723  & 0.8002  & 0.2086  & 0.0085  & 0.1980  & 0.2451  & 0.0469  & 0.2932  \\
    \bottomrule
    \end{tabular}%
  \label{tab:addlabel}%
\end{table*}%

To evaluate the robustness of the multi-step process under constrained supervision, we examine how varying training data sizes (100\%, 80\%, 60\%, 40\%, and 20\%) impact Qwen-7B’s performance across the SE lifecycle, with the test set fixed to isolate generalization effects.

As shown in Table~\ref{tab:3}, the pipeline remains stable even with substantial training data reduction. With only 20\% of the training data, most metrics remain close to the full-data setting. For instance, in code generation, CodeBLEU slightly improves from 0.2923 to 0.2932, indicating preserved or even enhanced syntactic and semantic fidelity. In detailed design, EM changes minimally from 0.2055 (full) to 0.2086 (20\%), with BLEU and ROUGE also showing minor variation (e.g., ROUGE: 0.1965 vs. 0.1980).

Interestingly, moderate training data reduction (e.g., 60–80\%) occasionally yields performance gains. For example, the requirements analysis BLEU improves from 0.0582 (100\%) to 0.0642 (80\%) and 0.0635 (60\%), while the architectural design EM increases from 0.3457 (100\%) to 0.3688 (80\%) and 0.3675 (60\%). These observations suggest that smaller data subsets may reduce overfitting and encourage more generalized representations in some stages of the pipeline. 

The architectural design (SCXML generation) phase shows the strongest stability, with TF-IDF similarity increasing from 0.7745 (100\%) to 0.8002 (20\%), reflecting the model’s consistent ability to preserve symbolic structure and control logic. Similarly, requirements and detailed design phases maintain high content fidelity, with only slight fluctuations in ROUGE and EM.

Overall, these results highlight the robustness of our pipeline-level supervision framework: multi-stage reasoning and intermediate supervision not only improve output quality but also confer resilience under limited data regimes. This robustness under low-resource conditions makes the pipeline approach especially attractive for domains where labeled data is scarce but high-quality modular priors (e.g., specifications, designs) are available. 

\subsection{RQ5: How does each intermediate artifact contribute to final code quality?}
\begin{figure}[h]
  \centering
  \includegraphics[width=\linewidth]{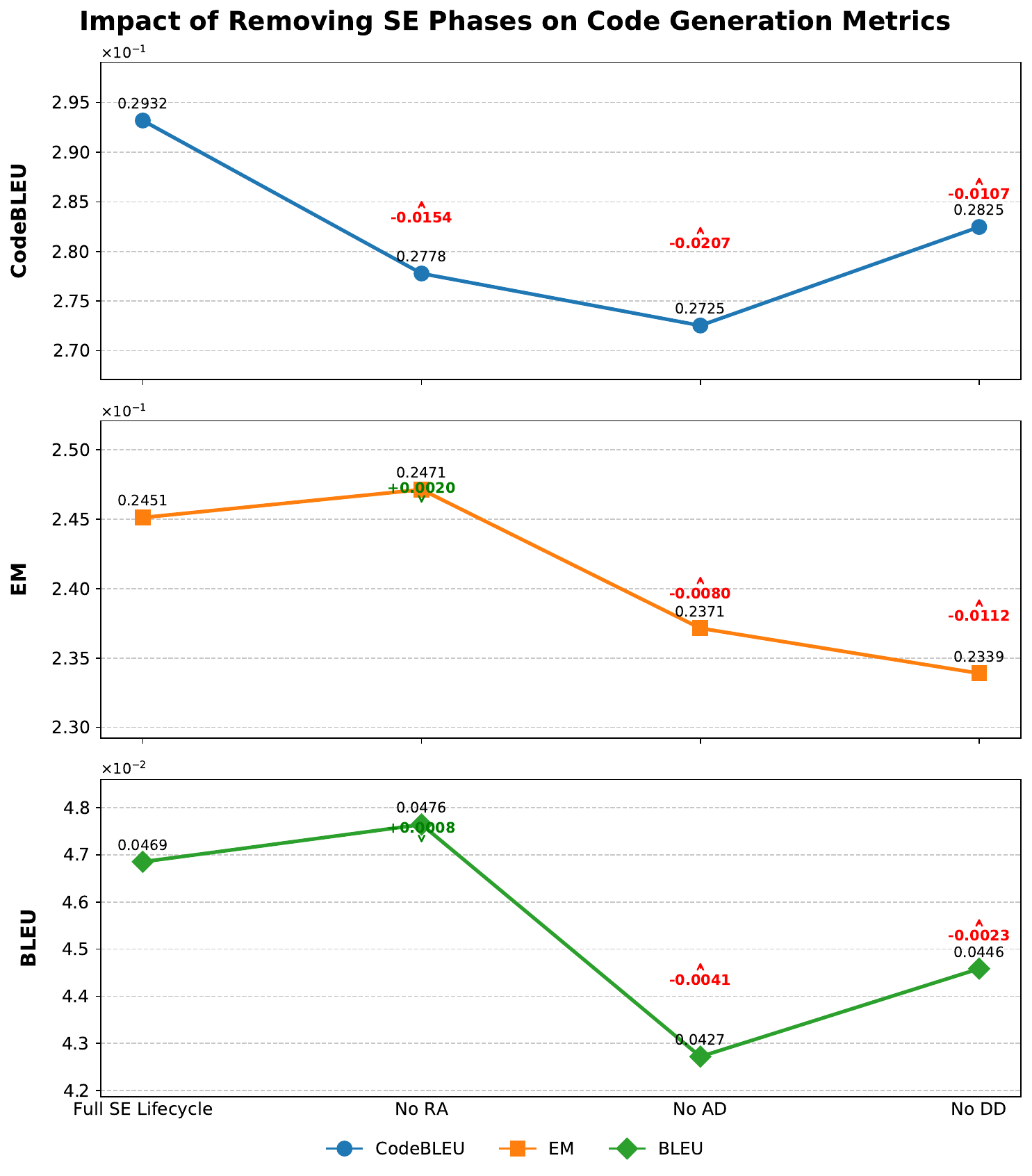}
  \caption{Contribution of each intermediate artifact to final code quality under ablation, using Qwen2.5-7B.}
  \label{fig:5}
\end{figure}

To assess the contribution of each intermediate artifact in the SE lifecycle, we conduct an ablation study using Qwen-7B with 20\% of the training data. We compare full lifecycle fine-tuning with three reduced variants, each removing one stage—requirements analysis, architectural design, or pseudocode generation—while keeping input-output mappings intact by substituting adjacent-stage data.

 As shown in Fig.~\ref{fig:5}, removing any single stage leads to a drop in CodeBLEU, confirming the importance of full SE supervision. EM and BLEU vary slightly but without a clear trend.

When the requirements analysis stage is omitted, and the model is trained using human intent–SCXML pairs in place of the human intent–detailed requirement pairs, CodeBLEU drops to 0.2778 (–5.3\%), and EM slightly increases to 0.2471, while BLEU improves marginally to 0.0476. This suggests that although early-stage supervision does not always reflect in exact match or token-level scores, it contributes to deeper structural quality captured by CodeBLEU. The absence of detailed requirements may reduce the model’s ability to disambiguate intent, leading to degraded semantic fidelity in the final code.

Omitting architectural design leads to the more substantial performance drop. Without SCXML, and using requirement–pseudocode pairs, CodeBLEU falls to 0.2725 (–7.1\%), EM to 0.2371, and BLEU to 0.0427. These results indicate that structured behavioral specifications (e.g., control flow and transitions) provided by SCXML are critical for encoding executable logic. Without explicit control design, the model struggles to infer precise procedural behavior, resulting in both lower lexical and semantic performance.

Removing pseudocode generation results in CodeBLEU of 0.2825 (–3.7\%), EM of 0.2339 (–4.5\%), and BLEU of 0.0446. Though the decline is smaller, it is consistent across metrics. The EM drop suggests that pseudocode plays a unique role in bridging symbolic specifications with executable syntax. By omitting this abstraction layer, the model’s ability to align functional structure with syntactic form is weakened, leading to reduced literal correctness.

Overall, all intermediate stages contribute to final performance. Architectural design has the most significant impact due to its formal control logic, followed by requirements analysis and pseudocode. These results support the effectiveness of our lifecycle-aware supervision, highlighting the complementary roles of natural language understanding, structured modeling, and algorithmic abstraction in improving code generation.

\section{Discussion}

In this section, we discuss the broader implications of our SE lifecycle-aware code generation framework, interpret the results, examine limitations, and identify potential threats to the validity of our findings.

\subsection{Interpretation of Results}

ur results show that integrating SE lifecycle principles into the code generation process significantly enhances both functional correctness and maintainability. The staged approach—from requirements to architecture, detailed design, and final code—produces interpretable artifacts that improve control, support fine-grained adjustments, and reduce prompt ambiguity and hallucinations.

\subsection{Framework Benefits and Limitations}

The proposed framework offers several practical advantages. By enforcing a structured design process that closely mirrors how human developers typically approach software construction, it enhances both logical consistency and development rigor. Its use of traceable intermediate representations improves explainability, allowing developers to better understand and verify the generation process. Furthermore, the modular organization of the pipeline, particularly the separation between detailed design and final code generation, enables straightforward adaptation to multiple programming languages by adjusting the translator component.

Nonetheless, the framework is not without limitations. The multi-stage architecture inevitably introduces latency, which may hinder its applicability in real-time or low-latency development scenarios. Moreover, since the stages are interdependent, errors introduced in early phases such as requirements analysis can propagate and amplify through the subsequent steps, potentially degrading the quality of the final output.

\subsection{Threats to Validity}

We categorize threats to validity into two types:

\textbf{Internal Validity:} Our model is fine-tuned end-to-end using a unified dataset that captures all four stages of the generation pipeline within each training instance. This approach enables the model to jointly learn the dependencies across phases, but also introduces the risk that errors in one part of the output (e.g., requirements analysis) may influence others (e.g., pseudocode quality). Since we do not train each stage independently, it becomes more difficult to isolate the source of specific generation errors. We attempt to mitigate this by ensuring high-quality alignment of all intermediate representations within each training tuple.

\textbf{External Validity:} Our evaluation is based on a curated corpus derived from real-world Aviation standards, commercial software documentation, and open-source projects. While these sources provide industrial relevance and domain diversity, the generalizability of our findings to entirely unrelated domains (such as video game development) may still be limited. Further studies are needed to test the adaptability of our approach across additional application domains and programming paradigms.

\subsection{Future Work}

Future efforts could explore integrating automated verification tools within each generation stage to detect specification mismatches early. Another promising direction is fine-tuning LLMs specifically for each phase of the lifecycle, potentially improving coherence and specialization. Additionally, exploring multilingual code generation pipelines may expand the applicability of the framework to non-English programming environments. Furthermore, exploring the application of RAG technology within the framework may better meet the professional requirements for the design of complex systems in specific fields.

\section{Conclusion}

This paper introduces SE lifecycle-aware code generation, which integrates software engineering phases into LLM-based code generation. By dividing the process into requirement analysis, architectural design, detailed design, and code generation, we combine the rigor of traditional SE with the power of LLMs, ensuring traceable and verifiable results.

We curated a dataset aligned with the RTCA/DO-185B standard and industrial practices, allowing fine-tuning of models as small as 1.3B parameters to outperform larger baselines. Our evaluation shows:
\begin{itemize}
    \item Fine-tuning with SE lifecycle data improves correctness by up to 75\%.
    \item Multi-step inference outperforms single-step generation.
    \item LLMs fine-tuned with the SE lifecycle perform as well as or better than code-pretrained models.
    \item The pipeline remains effective with 80\% less training data.
    \item Each intermediate artifact, particularly architectural design, improves final code quality.
\end{itemize}

These results create a new paradigm where LLMs are guided by engineering discipline, ensuring trustworthiness in automated software development. Our modular, model-agnostic framework is extensible and we release all resources for reproducibility and future research.

Future work will focus on expanding architectural paradigms, integrating formal verification for certified code generation, and exploring human-in-the-loop refinement. Ultimately, we envision LLMs as partners in the software development lifecycle, not just code generators.

\bibliographystyle{unsrt}
\bibliography{sample-base}

\begin{thebibliography}{10}

\bibitem{chen2021evaluatinglargelanguagemodels}
Mark Chen, Jerry Tworek, Heewoo Jun, Qiming Yuan, Henrique~Ponde
  de~Oliveira~Pinto, Jared Kaplan, Harri Edwards, Yuri Burda, Nicholas Joseph,
  Greg Brockman, Alex Ray, Raul Puri, Gretchen Krueger, Michael Petrov, Heidy
  Khlaaf, Girish Sastry, Pamela Mishkin, Brooke Chan, Scott Gray, Nick Ryder,
  Mikhail Pavlov, Alethea Power, Lukasz Kaiser, Mohammad Bavarian, Clemens
  Winter, Philippe Tillet, Felipe~Petroski Such, Dave Cummings, Matthias
  Plappert, Fotios Chantzis, Elizabeth Barnes, Ariel Herbert-Voss,
  William~Hebgen Guss, Alex Nichol, Alex Paino, Nikolas Tezak, Jie Tang, Igor
  Babuschkin, Suchir Balaji, Shantanu Jain, William Saunders, Christopher
  Hesse, Andrew~N. Carr, Jan Leike, Josh Achiam, Vedant Misra, Evan Morikawa,
  Alec Radford, Matthew Knight, Miles Brundage, Mira Murati, Katie Mayer, Peter
  Welinder, Bob McGrew, Dario Amodei, Sam McCandlish, Ilya Sutskever, and
  Wojciech Zaremba.
\newblock Evaluating large language models trained on code, 2021.

\bibitem{li2022competitionlevelcodegenerationalphacode}
Yujia Li, David Choi, Junyoung Chung, Nate Kushman, Julian Schrittwieser, Rémi
  Leblond, Tom Eccles, James Keeling, Felix Gimeno, Agustin~Dal Lago, Thomas
  Hubert, Peter Choy, Cyprien de~Masson~d'Autume, Igor Babuschkin, Xinyun Chen,
  Po-Sen Huang, Johannes Welbl, Sven Gowal, Alexey Cherepanov, James Molloy,
  Daniel~J. Mankowitz, Esme~Sutherland Robson, Pushmeet Kohli, Nando
  de~Freitas, Koray Kavukcuoglu, and Oriol Vinyals.
\newblock Competition-level code generation with alphacode, 2022.

\bibitem{chen2021evaluating}
Mark Chen, Jerry Tworek, Heewoo Jun, Qiming Yuan, Henrique Ponde de~Oliveira
  Pinto, Jared Kaplan, Harri Edwards, Yuri Burda, Nicholas Joseph, Greg
  Brockman, Alex Ray, Raul Puri, Gretchen Krueger, Michael Petrov, Heidy
  Khlaaf, Girish Sastry, Pamela Mishkin, Brooke Chan, Scott Gray, Nick Ryder,
  Mikhail Pavlov, Alethea Power, Matthias Plappert, Fotios Chantzis, Elizabeth
  Barnes, Ariel Herbert-Voss, William Hebgen~Guss, Alex Nichol, Alex Paino,
  Nikolas Tezak, Jie Tang, Igor Babuschkin, Suchir Balaji, Shantanu Jain,
  William Saunders, Christopher Hesse, Andrew~N. Carr, Jan Leike, Josh Achiam,
  Vedant Misra, Evan Morikawa, Alec Radford, Matthew Knight, Miles Brundage,
  Mira Murati, Katie Mayer, Peter Welinder, Bob McGrew, Dario Amodei, Sam
  McCandlish, Ilya Sutskever, and Wojciech Zaremba.
\newblock Evaluating large language models trained on code.
\newblock {\em arXiv preprint arXiv:2107.03374}, 2021.

\bibitem{hendrycks2021measuring}
Dan Hendrycks, Steven Basart, Saurav Kadavath, Mantas Mazeika, Akul Arora,
  Ethan Guo, Collin Burns, Samir Puranik, Horace He, Dawn Song, and Jacob
  Steinhardt.
\newblock Measuring coding challenge competence with apps.
\newblock {\em arXiv preprint arXiv:2105.09938}, 2021.

\bibitem{brooks1995mythical}
Frederick~P. Brooks.
\newblock {\em The Mythical Man-Month}.
\newblock Addison-Wesley, 1995.

\bibitem{royce1970management}
Winston~W. Royce.
\newblock Managing the development of large software systems.
\newblock In {\em Proceedings of IEEE WESCON}, pages 1--9, Los Angeles, CA,
  1970.

\bibitem{somerville2016software}
Ian Somerville.
\newblock {\em Software Engineering}.
\newblock Pearson, 10 edition, 2016.

\bibitem{lehtinen2014perceived}
Timo O.~A. Lehtinen, Mika~V. Mäntylä, Jari Vanhanen, Juha Itkonen, and Casper
  Lassenius.
\newblock Perceived causes of software project failures: An analysis of their
  relationships.
\newblock {\em Information and Software Technology}, 56(6):623--643, 2014.

\bibitem{perry1989foundations}
Dewayne~E. Perry and Alexander~L. Wolf.
\newblock Foundations for the study of software architecture.
\newblock {\em ACM SIGSOFT Software Engineering Notes}, 17(4):40--52, 1989.

\bibitem{chen2023evaluating}
Zhiqiang Yuan, Yiling Lou, Mingwei Liu, Shiji Ding, Kaixin Wang, Yixuan Chen,
  and Xin Peng.
\newblock No more manual tests? evaluating and improving chatgpt for unit test
  generation.
\newblock {\em ArXiv}, 2023.

\bibitem{zhang2023repocoder}
Fengji Zhang, Bei Chen, Yue Zhang, Jacky Keung, Jin Liu, Daoguang Zan, Yi~Mao,
  Jian‑Guang Lou, and Weizhu Chen.
\newblock Repocoder: Repository‑level code completion through iterative
  retrieval and generation.
\newblock In {\em Proceedings of EMNLP}, pages 2470--2481. ACL, 2023.

\bibitem{barnett2015state}
Jim Barnett, Rahul Akolkar, RJ~Auburn, Michael Bodell, Daniel~C. Burnett, Jerry
  Carter, Scott McGlashan, Torbjörn Lager, Mark Helbing, Rafah Hosn, T.V.
  Raman, Klaus Reifenrath, No'am Rosenthal, and Johan Roxendal.
\newblock State chart xml (scxml): State machine notation for control
  abstraction.
\newblock Technical report, Massachusetts, USA, 2015.

\bibitem{jimenez2024evaluation}
{\'A}lvaro~Barbero Jim{\'e}nez.
\newblock An evaluation of llm code generation capabilities through graded
  exercises.
\newblock {\em arXiv preprint arXiv:2410.16292}, 2024.

\bibitem{ouyang2023llm}
Shuyin Ouyang, Jie~M Zhang, Mark Harman, and Meng Wang.
\newblock Llm is like a box of chocolates: the non-determinism of chatgpt in
  code generation.
\newblock {\em arXiv e-prints}, pages arXiv--2308, 2023.

\bibitem{donato2025studying}
Benedetta Donato, Leonardo Mariani, Daniela Micucci, and Oliviero Riganelli.
\newblock Studying how configurations impact code generation in llms: The case
  of chatgpt.
\newblock {\em arXiv preprint arXiv:2502.17450}, 2025.

\bibitem{liu2024exploring}
Fang Liu, Yang Liu, Lin Shi, Houkun Huang, Ruifeng Wang, Zhen Yang, Li~Zhang,
  Zhongqi Li, and Yuchi Ma.
\newblock Exploring and evaluating hallucinations in llm-powered code
  generation.
\newblock {\em arXiv preprint arXiv:2404.00971}, 2024.

\bibitem{liu2025code}
Mingwei Liu, Juntao Li, Ying Wang, Xueying Du, Zuoyu Ou, Qiuyuan Chen, Bingxu
  An, Zhao Wei, Yong Xu, Fangming Zou, et~al.
\newblock Code copycat conundrum: Demystifying repetition in llm-based code
  generation.
\newblock {\em arXiv preprint arXiv:2504.12608}, 2025.

\bibitem{rahman2025large}
Musfiqur Rahman, SayedHassan Khatoonabadi, and Emad Shihab.
\newblock A large-scale class-level benchmark dataset for code generation with
  llms.
\newblock {\em arXiv preprint arXiv:2504.15564}, 2025.

\bibitem{murr2023testing}
Lincoln Murr, Morgan Grainger, and David Gao.
\newblock Testing llms on code generation with varying levels of prompt
  specificity.
\newblock {\em arXiv preprint arXiv:2311.07599}, 2023.

\bibitem{liu2023pre}
Pengfei Liu, Weizhe Yuan, Jinlan Fu, Zhengbao Jiang, Hiroaki Hayashi, and
  Graham Neubig.
\newblock Pre-train, prompt, and predict: A systematic survey of prompting
  methods in natural language processing.
\newblock {\em ACM computing surveys}, 55(9):1--35, 2023.

\bibitem{zhang2024renaissance}
Wuyang Zhang, Yansong Li, Zeyu Dong, Yu~Wu, Yingyao Zhou, Duolei Wang,
  Songsirou Xing, Chichun Zhou, and Da~Shen.
\newblock Renaissance of literate programming in the era of llms: Enhancing
  llm-based code generation in large-scale projects.
\newblock {\em arXiv preprint arXiv:2502.17441}, 2024.

\bibitem{knuth1984literate}
Donald~Ervin Knuth.
\newblock Literate programming.
\newblock {\em The computer journal}, 27(2):97--111, 1984.

\bibitem{assogbaevaluating}
Yannick Assogba and Donghao Ren.
\newblock Evaluating long range dependency handling in code generation llms.
\newblock {\em Transactions on Machine Learning Research}.

\bibitem{wang2025leveraging}
Shuai Wang, Liang Ding, Yibing Zhan, Yong Luo, Zheng He, and Dapeng Tao.
\newblock Leveraging metamemory mechanisms for enhanced data-free code
  generation in llms.
\newblock {\em arXiv preprint arXiv:2501.07892}, 2025.

\bibitem{busch2025llm}
Daniel Busch, Alexander Bainczyk, Steven Smyth, and Bernhard Steffen.
\newblock Llm-based code generation and system migration in language-driven
  engineering.
\newblock {\em International Journal on Software Tools for Technology
  Transfer}, 27(1):137--147, 2025.

\bibitem{wang2023chatcoder}
Zejun Wang, Jia Li, Ge~Li, and Zhi Jin.
\newblock Chatcoder: Chat-based refine requirement improves llms' code
  generation.
\newblock {\em arXiv preprint arXiv:2311.00272}, 2023.

\bibitem{dutta2024applying}
Sujan Dutta, Sayantan Mahinder, Raviteja Anantha, and Bortik Bandyopadhyay.
\newblock Applying rlaif for code generation with api-usage in lightweight
  llms.
\newblock {\em arXiv preprint arXiv:2406.20060}, 2024.

\bibitem{sit2024developing}
Chun Yan~Enoch SIT, YANG Yin, Wing~Kei YEUNG, and Siu~Cheung KONG.
\newblock Developing a llms-driven system based on human-ai progressive code
  generation framework to assist mathematics learning.
\newblock In {\em International Conference on Computers in Education}, 2024.

\bibitem{mu2024clarifygpt}
Fangwen Mu, Lin Shi, Song Wang, Zhuohao Yu, Binquan Zhang, ChenXue Wang,
  Shichao Liu, and Qing Wang.
\newblock Clarifygpt: A framework for enhancing llm-based code generation via
  requirements clarification.
\newblock {\em Proceedings of the ACM on Software Engineering},
  1(FSE):2332--2354, 2024.

\bibitem{ashrafi2025enhancing}
Nazmus Ashrafi, Salah Bouktif, and Mohammed Mediani.
\newblock Enhancing llm code generation: A systematic evaluation of multi-agent
  collaboration and runtime debugging for improved accuracy, reliability, and
  latency.
\newblock {\em arXiv preprint arXiv:2505.02133}, 2025.

\bibitem{xie2025empowering}
Chen Xie, Mingsheng Jiao, Xiaodong Gu, and Beijun Shen.
\newblock Empowering ai to generate better ai code: Guided generation of deep
  learning projects with llms.
\newblock {\em arXiv preprint arXiv:2504.15080}, 2025.

\bibitem{zhu2025uncertainty}
Yuqi Zhu, Ge~Li, Xue Jiang, Jia Li, Hong Mei, Zhi Jin, and Yihong Dong.
\newblock Uncertainty-guided chain-of-thought for code generation with llms.
\newblock {\em arXiv preprint arXiv:2503.15341}, 2025.

\bibitem{wei2022chain}
Jason Wei, Xuezhi Wang, Dale Schuurmans, Maarten Bosma, Fei Xia, Ed~Chi, Quoc~V
  Le, Denny Zhou, et~al.
\newblock Chain-of-thought prompting elicits reasoning in large language
  models.
\newblock {\em Advances in neural information processing systems},
  35:24824--24837, 2022.

\bibitem{tao2024codelutra}
Leitian Tao, Xiang Chen, Tong Yu, Tung Mai, Ryan Rossi, Yixuan Li, and Saayan
  Mitra.
\newblock Codelutra: Boosting llm code generation via preference-guided
  refinement.
\newblock {\em arXiv preprint arXiv:2411.05199}, 2024.

\bibitem{zuo2025complexvcoder}
Jian Zuo, Junzhe Liu, Xianyong Wang, Yicheng Liu, Navya Goli, Tong Xu, Hao
  Zhang, Umamaheswara~Rao Tida, Zhenge Jia, and Mengying Zhao.
\newblock Complexvcoder: An llm-driven framework for systematic generation of
  complex verilog code.
\newblock {\em arXiv preprint arXiv:2504.20653}, 2025.

\bibitem{vijayaraghavan2024chain}
Prashanth Vijayaraghavan, Apoorva Nitsure, Charles Mackin, Luyao Shi, Stefano
  Ambrogio, Arvind Haran, Viresh Paruthi, Ali Elzein, Dan Coops, David Beymer,
  et~al.
\newblock Chain-of-descriptions: Improving code llms for vhdl code generation
  and summarization.
\newblock In {\em Proceedings of the 2024 ACM/IEEE International Symposium on
  Machine Learning for CAD}, pages 1--10, 2024.

\bibitem{murphy2024combining}
William Murphy, Nikolaus Holzer, Feitong Qiao, Leyi Cui, Raven Rothkopf, Nathan
  Koenig, and Mark Santolucito.
\newblock Combining llm code generation with formal specifications and reactive
  program synthesis.
\newblock {\em arXiv preprint arXiv:2410.19736}, 2024.

\bibitem{zhao2024empowering}
Yang Zhao, Di~Huang, Chongxiao Li, Pengwei Jin, Ziyuan Nan, Tianyun Ma, Lei Qi,
  Yansong Pan, Zhenxing Zhang, Rui Zhang, et~al.
\newblock Empowering llms for verilog generation through multi-level
  summarization.
\newblock {\em arXiv e-prints}, pages arXiv--2407, 2024.

\bibitem{yaacov2024boosting}
Tom Yaacov, Achiya Elyasaf, and Gera Weiss.
\newblock Boosting llm-based software generation by aligning code with
  requirements.
\newblock In {\em 2024 IEEE 32nd International Requirements Engineering
  Conference Workshops (REW)}, pages 301--305. IEEE, 2024.

\bibitem{almorsi2025guided}
Amr Almorsi, Mohanned Ahmed, and Walid Gomaa.
\newblock Guided code generation with llms: A multi-agent framework for complex
  code tasks.
\newblock {\em arXiv preprint arXiv:2501.06625}, 2025.

\bibitem{liu2024agents4plc}
Zihan Liu, Ruinan Zeng, Dongxia Wang, Gengyun Peng, Jingyi Wang, Qiang Liu,
  Peiyu Liu, and Wenhai Wang.
\newblock Agents4plc: Automating closed-loop plc code generation and
  verification in industrial control systems using llm-based agents.
\newblock {\em arXiv preprint arXiv:2410.14209}, 2024.

\bibitem{lin2024llm}
Feng Lin, Dong~Jae Kim, and Tse-Hsun Chen.
\newblock When llm-based code generation meets the software development
  process.
\newblock {\em arXiv preprint arXiv:2403.15852}, 2024.

\bibitem{mahmud2025enhancing}
Tarek Mahmud, Bin Duan, Corina Pasareanu, and Guowei Yang.
\newblock Enhancing llm code generation with ensembles: A similarity-based
  selection approach.
\newblock {\em arXiv preprint arXiv:2503.15838}, 2025.

\bibitem{ren2020codebleu}
Shuo Ren, Daya Guo, Shuai Lu, Long Zhou, Shujie Liu, Duyu Tang, Neel
  Sundaresan, Ming Zhou, Ambrosio Blanco, and Shuai Ma.
\newblock Codebleu: a method for automatic evaluation of code synthesis.
\newblock {\em arXiv preprint arXiv:2009.10297}, 2020.

\bibitem{pan2025codecor}
Ruwei Pan, Hongyu Zhang, and Chao Liu.
\newblock Codecor: An llm-based self-reflective multi-agent framework for code
  generation.
\newblock {\em arXiv preprint arXiv:2501.07811}, 2025.

\bibitem{lin2025autop2c}
Zijie Lin, Yiqing Shen, Qilin Cai, He~Sun, Jinrui Zhou, and Mingjun Xiao.
\newblock Autop2c: An llm-based agent framework for code repository generation
  from multimodal content in academic papers.
\newblock {\em arXiv preprint arXiv:2504.20115}, 2025.

\bibitem{rtca2009do185b}
Minimum operational performance standards (mops) for airborne automatic
  dependent surveillance - broadcast (ads-b).
\newblock Technical Report DO-185B, RTCA, Inc., December 2009.
\newblock RTCA/DO-185B.

\bibitem{hu2022lora}
Edward~J Hu, Yelong Shen, Phillip Wallis, Zeyuan Allen-Zhu, Yuanzhi Li, Shean
  Wang, Lu~Wang, Weizhu Chen, et~al.
\newblock Lora: Low-rank adaptation of large language models.
\newblock {\em ICLR}, 1(2):3, 2022.

\bibitem{harel1987statecharts}
David Harel.
\newblock Statecharts: A visual formalism for complex systems.
\newblock {\em Science of Computer Programming}, 8(3):231--274, Jun 1987.

\bibitem{roziere2023code}
Baptiste Roziere, Jonas Gehring, Fabian Gloeckle, Sten Sootla, Itai Gat,
  Xiaoqing~Ellen Tan, Yossi Adi, Jingyu Liu, Romain Sauvestre, Tal Remez,
  et~al.
\newblock Code llama: Open foundation models for code.
\newblock {\em arXiv preprint arXiv:2308.12950}, 2023.

\bibitem{papineni2002bleu}
Kishore Papineni, Salim Roukos, Todd Ward, and Wei-Jing Zhu.
\newblock Bleu: a method for automatic evaluation of machine translation.
\newblock In {\em Proc. ACL}, 2002.

\bibitem{lin2004rouge}
Chin-Yew Lin.
\newblock Rouge: A package for automatic evaluation of summaries.
\newblock In {\em Text summarization branches out: Proc. ACL Workshop}, 2004.

\bibitem{salton1988term}
Gerard Salton and Christopher Buckley.
\newblock Term-weighting approaches in automatic text retrieval.
\newblock In {\em Information processing \& management}, volume~24, pages
  513--523. Elsevier, 1988.

\bibitem{team2024qwen2}
Qwen Team.
\newblock Qwen2 technical report.
\newblock {\em arXiv preprint arXiv:2407.10671}, 2024.

\bibitem{guo2024deepseek}
Daya Guo, Qihao Zhu, Dejian Yang, Zhenda Xie, Kai Dong, Wentao Zhang, Guanting
  Chen, Xiao Bi, Yu~Wu, YK~Li, et~al.
\newblock Deepseek-coder: When the large language model meets programming--the
  rise of code intelligence.
\newblock {\em arXiv preprint arXiv:2401.14196}, 2024.

\bibitem{achiam2023gpt}
Josh Achiam, Steven Adler, Sandhini Agarwal, Lama Ahmad, Ilge Akkaya,
  Florencia~Leoni Aleman, Diogo Almeida, Janko Altenschmidt, Sam Altman,
  Shyamal Anadkat, et~al.
\newblock Gpt-4 technical report.
\newblock {\em arXiv preprint arXiv:2303.08774}, 2023.

\bibitem{brown2020language}
Tom Brown, Benjamin Mann, Nick Ryder, Melanie Subbiah, Jared~D Kaplan, Prafulla
  Dhariwal, Arvind Neelakantan, Pranav Shyam, Girish Sastry, Amanda Askell,
  et~al.
\newblock Language models are few-shot learners.
\newblock {\em Advances in neural information processing systems},
  33:1877--1901, 2020.

\bibitem{guo2025deepseek}
Daya Guo, Dejian Yang, Haowei Zhang, Junxiao Song, Ruoyu Zhang, Runxin Xu,
  Qihao Zhu, Shirong Ma, Peiyi Wang, Xiao Bi, et~al.
\newblock Deepseek-r1: Incentivizing reasoning capability in llms via
  reinforcement learning.
\newblock {\em arXiv preprint arXiv:2501.12948}, 2025.

\bibitem{touvron2023llama}
Hugo Touvron, Thibaut Lavril, Gautier Izacard, Xavier Martinet, Marie-Anne
  Lachaux, Timoth{\'e}e Lacroix, Baptiste Rozi{\`e}re, Naman Goyal, Eric
  Hambro, Faisal Azhar, et~al.
\newblock Llama: Open and efficient foundation language models.
\newblock {\em arXiv preprint arXiv:2302.13971}, 2023.

\end{thebibliography}

\appendix

\end{sloppypar}
\end{document}


\appendix
\section{Prompts}
Table \ref{tab:pset} lists all the prompts used in the datasets and experiments.

\begin{center}
\begin{longtable}{| p{12cm} |}

\hline
\textbf{Prompt} \\
\hline
\endfirsthead

\hline
\endfoot

\textbf{Dataset and Multi-step Generation} \\
\hline
\texttt{INSTRUCTION: Generate a detailed requirement description based on the following original requirement description.\newline
  INPUT: [Initial Requirement] \newline
  OUTPUT: [Detailed Requirement]} \\
\hline
\texttt{INSTRUCTION: Generate a state machine design description in SCXML format based on the following detailed requirement description.\newline
  INPUT: [Detailed Requirement] \newline
  OUTPUT: [SCXML]} \\
\hline
\texttt{INSTRUCTION: Generate a pseudocode based on the following state machine design description in SCXML format.\newline
  INPUT: [SCXML] \newline
  OUTPUT: [Pseudocode]} \\
\hline
\texttt{INSTRUCTION: Generate an executable python program based on the following pseudocode.\newline
  INPUT: [Pseudocode] \newline
  OUTPUT: [Python Code]} \\
\hline

\textbf{One-step Generation} \\
\hline
\texttt{INSTRUCTION: Generate a detailed requirement description based on the following original requirement description.\newline
  INPUT: [Initial Requirement] \newline
  OUTPUT: [Detailed Requirement]} \\
\hline
\texttt{INSTRUCTION: Generate a state machine design description in SCXML format based on the following original requirement description.\newline
  INPUT: [Initial Requirement] \newline
  OUTPUT: [SCXML]} \\
\hline
\texttt{INSTRUCTION: Generate a pseudocode based on the following original requirement description.\newline
  INPUT: [Initial Requirement] \newline
  OUTPUT: [Pseudocode]} \\
\hline
\texttt{INSTRUCTION: Generate an executable python program based on the following original requirement description.\newline
  INPUT: [Initial Requirement] \newline
  OUTPUT: [Python Code]} \\
\hline

\textbf{Construct Dataset from Document}\\
    \hline
\texttt{INSTRUCTION: Please, from the perspective of a product manager, based on the following more detailed requirements description, provide a concise natural language requirement in one sentence. This requirement should exclude the specific state machine design and prohibit the inclusion of design-related descriptions in the requirement. Only simple functional requirements need to be given. Output in English. The detailed requirements description is as follows.\newline
  INPUT: [FSM Description] \newline
  OUTPUT: [Initial Requirement]}\\
  \hline
\texttt{INSTRUCTION: Based on the following design and description content, provide a complete requirement for this content, and output it in English. The detailed design and description content is as follows.\newline
  INPUT: [FSM Description] \newline
  OUTPUT: [Detailed Requirement]}\\
  \hline
\texttt{INSTRUCTION: You are an expert in the field of software engineering and are proficient in the software development process based on state machines. Please read carefully the given description of the state machine based on natural language, analyze the states and transition conditions therein, and convert it into the description of the state machine in SCXML format. The state machine based on natural language is described as follows.\newline
  INPUT: [FSM Description] \newline
  OUTPUT: [SCXML]}\\
  \hline
\texttt{INSTRUCTION: You are an expert in the field of software engineering. Please generate executable Python code based on the following pseudo code. Note that the comments in the Python code should be in English. The pseudo code is as follows.\newline
  INPUT: [Pseudocode] \newline
  OUTPUT: [Python Code]}\\
  \hline

\textbf{Construct Dataset from Open-source Code}\\
    \hline
\texttt{INSTRUCTION: You are an expert in the field of software engineering. I will give you a python program based on the state machine. Please read this program carefully and then summarize and describe it. The Description needs to include the functional description of this program, what states and transition conditions there are. Note that the output format should be "Description: functional description of the program, use an imperative sentence without a subject. States: describe the states of the state machine, mark the states by serial number. Transitions: describe under what conditions the state machine transition from which state to which state, mark the transitions by serial number."  Note that the output content must be a single piece of text and should not be output in segments. The python program you need to analyze is as follows.\newline
  INPUT: [Python Code] \newline
  OUTPUT: [FSM Description]}\\
  \hline
\texttt{INSTRUCTION: From the perspective of a product manager, please provide a brief natural language requirement in one sentence based on the following more detailed requirement description, use an imperative sentence without a subject. This requirement needs to shield the specific state machine design and prohibit the expression of design-related descriptions in the requirement. Only simple functional requirements need to be given. The detailed requirements are described as follows.\newline
  INPUT: [FSM Description] \newline
  OUTPUT: [Initial Requirement]}\\
  \hline
\texttt{INSTRUCTION: You are an expert in the field of software engineering. I will provide you with a state machine description. Please carefully analyze this requirement and produce a detailed functional description. Please note that the output format should be a single text, without paragraph breaks. The requirement description is as follows.\newline
  INPUT: [FSM Description] \newline
  OUTPUT: [Detailed Requirement]}\\
  \hline
\texttt{INSTRUCTION: You are an expert in the field of software engineering and are proficient in the software development process based on state machines. Please read carefully the given description of the state machine based on natural language, analyze the states and transition conditions therein, and convert it into the description of the state machine in SCXML format. The state machine based on natural language is described as follows.\newline
  INPUT: [FSM Description] \newline
  OUTPUT: [SCXML]}\\
  \hline
\texttt{INSTRUCTION: You are an expert in the field of software engineering. I will give you a python program. Please generate pseudo-code based on the provided python program. Meanwhile, I will give you a pseudo-code example. You must generate pseudo-code by imitating the provided pseudo-code example. The pseudo-code example is as follows: pseudocode sample, The python program is as follows.\newline
  INPUT: [Python Code] \newline
  OUTPUT: [Pseudocode]}\\
  \hline

\textbf{Construct Dataset from Seeds}\\
    \hline
\texttt{INSTRUCTION: You are an expert in the field of software engineering. Please generate a new set of data based on the sample of the provided code-generation task data according to the following requirements. Note that the generated data should be in JSON format. Data format requirements: The data contains the following five fields: 1. The "raw" field describes the core functions of the system in 1-2 sentences; 2. The "detail" field expands the requirements based on the original requirement description in the "raw" field. It needs to include the complete title, clear purpose, definitions of key terms, structured requirement list, system behavior description, performance requirements, test verification requirements, etc. Note that the format should be a separate paragraph of text and not segmented. 3. The "fsm" field conducts the design based on the state machine according to the detailed requirement description in the "detail" field. It needs to include the state machine description, state definition, transition conditions, trigger mechanism, etc. Pay attention to using natural language and logical operators for description, and the format is a separate paragraph of text. Do not divide into sections. 4. The "pseudocode" field generates pseudocode based on the state machine design of the "fsm" field and needs to include contents such as modular structure, input and output definitions, key algorithm logic, and industry standard naming conventions; 5. The "code" field generates python code based on the pseudo-code in the "pseudocode" field, which should include complete function definitions, input and output definitions, parameter validation, and code logic that conforms to the python specification, etc. Generation rule requirements: 1. Please read the content of the "pseudocode" field carefully, learn the format of pseudocode, and the generated pseudocode must conform to this format. 2. It must be a brand-new scene, and different fields can be selected (avionics systems, medical equipment control, industrial automation, intelligent transportation systems, energy management systems, game development, robotics technology, hardware design, network protocols, compiler design, user interface design, etc.), including but not limited to the above fields. 3. Strict technical consistency must be maintained at all levels. 4. The mapping from requirements to design is traceable. 5. The pseudo-code needs to have a logical correspondence with the python implementation. 6. Incorporate appropriate technical complexity (such as handling boundary conditions and exceptional situations). The code generated data sample is as follows.\newline
  INPUT: [Seed Data] \newline
  OUTPUT: [Evolved Data]}\\
  \hline

\caption{Complete set of prompts.}
\label{tab:pset}

\end{longtable}
\end{center}